\documentclass[runningheads]{llncs}
\usepackage{mathrsfs}
\usepackage{amsmath}
\usepackage{amssymb}

\usepackage{color}
\usepackage{xcolor}

\usepackage{graphicx}

\usepackage{url}
\newcommand{\doi}[1]{\textsc{doi}: \href{http://dx.doi.org/#1}{\nolinkurl{#1}}}

\makeatletter
\RequirePackage[bookmarks,unicode,colorlinks=true]{hyperref}%
   \def\@citecolor{blue}%
   \def\@urlcolor{blue}%
   \def\@linkcolor{blue}%

\def\orcidID#1{\smash{\href{http://orcid.org/#1}{\protect\raisebox{-1.25pt}{\protect\includegraphics{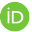}}}}}
\makeatother

\usepackage{hyperref}

\newcommand{\Adam}{\textsc{Adam}}
\newcommand{\AdamMC}{\textsc{AdamMC}}
\newcommand{\AdamSYNT}{\textsc{AdamSYNT}}

\newcommand{\refFig}[1]{Fig.~\ref{fig:#1}}
\newcommand{\pNet}{\ensuremath\mathcal{N}}

\usepackage[firstpage]{draftwatermark}
\SetWatermarkText{\hspace*{6in}\raisebox{6.3in}{\includegraphics[scale=0.1]{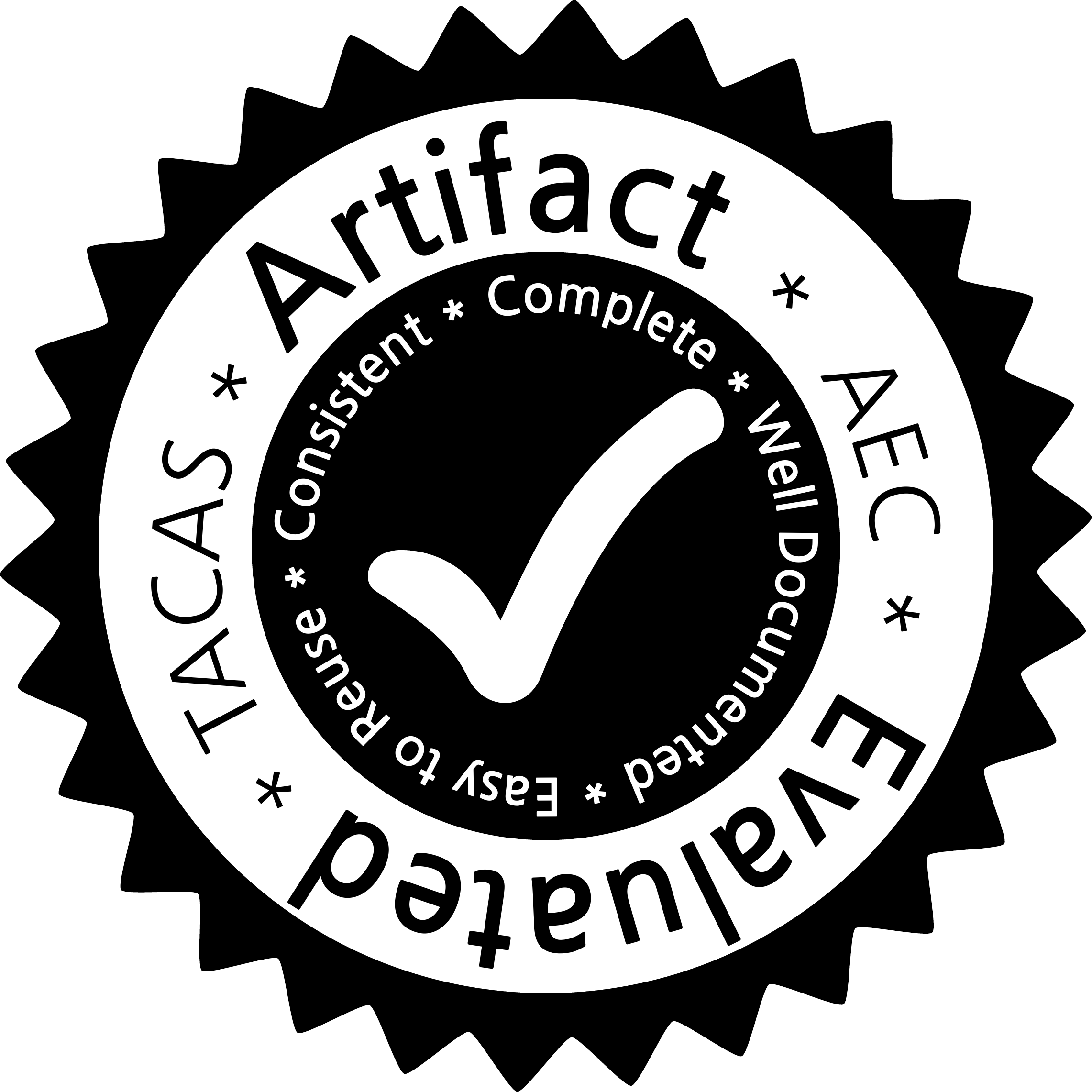}}}
\SetWatermarkAngle{0}

\usepackage{marvosym}

\usepackage[stable]{footmisc}

\begin{document}
\title{A Web Interface for Petri Nets with Transits and Petri Games\ %
\footnote[3]{This is the authors' copy of \cite{DBLP:conf/tacas/GiesekingHY21}.}%
\thanks{This work has been supported by the German Research Foundation (DFG) through Grant Petri Games (392735815) and through the Collaborative Research Center ``Foundations of Perspicuous Software Systems'' (TRR 248, 389792660), and by the European Research Council (ERC) through Grant OSARES (683300).}%
}
\titlerunning{A Web Interface for Petri Nets with Transits and Petri Games}
%
\author{
Manuel Gieseking\inst{1}\textsuperscript{(\Letter)}\orcidID{0000-0001-9073-3002} \and
Jesko Hecking-Harbusch\inst{2}\orcidID{0000-0003-2076-617X} \and
Ann Yanich\inst{1}\orcidID{0000-0002-0170-0012}
}

\authorrunning{M.\ Gieseking et al.}
%
\institute{
University of Oldenburg, Oldenburg, Germany\\
\email{\{gieseking,ann.yanich\}@informatik.uni-oldenburg.de}
\and
CISPA Helmholtz Center for Information Security, Saarbr\"ucken, Germany\\
\email{jesko.hecking-harbusch@cispa.de}
}

\maketitle

\begin{abstract}
\setcounter{footnote}{0}
Developing algorithms for distributed systems is an error-prone task.
Formal models like Petri nets with transits and Petri games can prevent errors when developing such algorithms.
Petri nets with transits allow us to follow the data flow between components in a distributed system.
They can be model checked against specifications in LTL on both the local data flow and the global behavior.
Petri games allow the synthesis of local controllers for distributed systems from safety specifications.
Modeling problems in these formalisms requires defining extended Petri nets which can be cumbersome when performed textually.

In this paper, we present a web interface\footnote{The web interface is deployed at \url{http://adam.informatik.uni-oldenburg.de}.} that allows an intuitive, visual definition of Petri nets with transits and Petri games.
The corresponding model checking and synthesis problems are solved directly on a server.
In the interface, implementations, counterexamples, and all intermediate steps can be analyzed and simulated.
Stepwise simulations and interactive state space generation support the user in detecting modeling errors.
\end{abstract}

\section{Introduction}
\label{sec:intro}
Distributed systems consist of several individual components.
Each component has incomplete information about the other components.
Asynchronous distributed systems have no fixed rate at which components progress but rather each component progresses at its individual rate between synchronizations with other components.
Implementing correct algorithms for asynchronous distributed systems is difficult because they have to both work with the incomplete information of the components and for every possible scheduling between the components.

\emph{Petri nets}~\cite{DBLP:books/sp/Reisig85a,DBLP:journals/tcs/NielsenPW81} are a natural model for asynchronous distributed systems.
Tokens represent components and transitions with more than one token correspond to synchronizations between the components.
\emph{Petri nets with transits}~\cite{DBLP:conf/atva/FinkbeinerGHO19} extend Petri nets with a transit relation to model the data flow in asynchronous distributed systems.
\emph{Flow-LTL}~\cite{DBLP:conf/atva/FinkbeinerGHO19} is a specification language for Petri nets with transits and allows us to specify linear properties on both the global and the local view of the system.
In particular, it is possible to globally select desired runs of the system with LTL  (e.g., only fair and maximal runs) and check the local data flow of only those runs again with LTL.
A model checker for Petri nets with transits against Flow-LTL is implemented in the tool \AdamMC~\cite{DBLP:conf/cav/FinkbeinerGHO20}.

\emph{Petri games}~\cite{DBLP:journals/iandc/FinkbeinerO17} define the synthesis of asynchronous distributed systems based on Petri nets and causal memory.
With causal memory, players exchange their entire causal past only upon synchronization.
Without synchronization, players have no information of each other.
For safety winning conditions, the synthesis algorithm for Petri games with a bounded number of controllable components and one uncontrollable component is implemented in \AdamSYNT~\cite{DBLP:conf/cav/FinkbeinerGO15}\footnote{\AdamSYNT{} was previously called \Adam. From now on, \AdamMC{} and \AdamSYNT{} are combined in the tool \Adam{} (\href{https://github.com/adamtool/adam\#readme}{https://github.com/adamtool/adam}).}.
Both tools are command-line tools lacking visual support to model Petri nets with transits or Petri games and the possibility to simulate or interactively explore implementations, counterexamples, and parts of the created state space.

In this paper, we present a web interface\footnote{The web interface is open source (\href{https://github.com/adamtool/webinterface\#readme}{https://github.com/adamtool/webinterface})
and a corresponding artifact to set it all up locally in a virtual machine is available~\cite{GHY20}.} for model checking asynchronous distributed systems with data flows and for the synthesis of asynchronous distributed systems with causal memory from safety specification.
The web interface offers an input for Petri nets with transits and Petri games where the user interactively creates places, transitions, and their connections with a few inputs.

As a back-end, the algorithms of \AdamMC{} are used to model check Petri nets with transits against a given Flow-LTL formula as specification.
Internally, the problem is reduced to the model checking problem of Petri nets against LTL.
Both, the input Petri net with transits and the constructed Petri net can be visualized and simulated in the web interface.
For a positive result, the web interface lets the user follow the control flow of the combined system and the data flow of the components.
For a negative result, the web interface simulates the counterexample with a visual separation of the global and each local behavior.

The algorithms of \AdamSYNT{} solve the given Petri game with safety specification.
Internally, the problem is reduced to solving a finite two-player game with complete information.
For a positive result, a winning strategy for the Petri game and the two-player game can be visualized and
the former can be simulated.
For a negative result, the web interface lets the user interactively construct strategies of the two-player game and highlights why they violate the specification.
These new intuitive construction methods, interactive features, and visualizations are of great impact when developing asynchronous distributed systems.

\section{Web Interface for Petri Nets with Transits}
\label{sec:PNwT}

The web interface can model check Petri nets with transits against Flow-LTL.
We use an example from software-defined networks to showcase the workflow.

\subsubsection{Workflow for Petri Nets With Transits}
\label{sec:PNwTworkflow}

\begin{figure}[t]
	\centering
	\includegraphics[width=\textwidth]{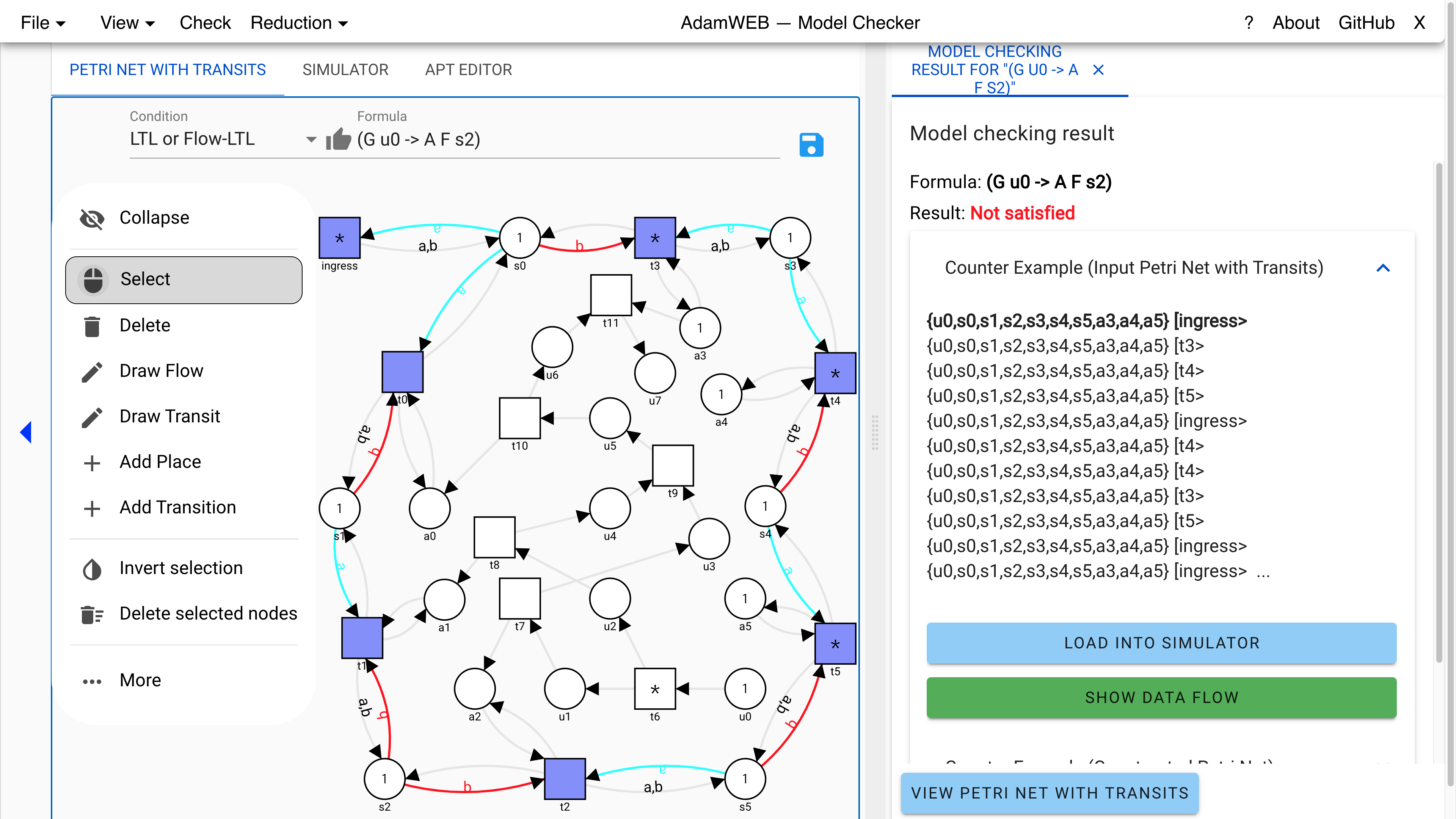}
	\caption{Screenshot from the web interface for the model checking workflow.}
	\label{fig:pnwt}
\end{figure}

One application domain for Petri nets with transits are \emph{software-defined networks (SDNs)}~\cite{DBLP:journals/ccr/McKeownABPPRST08,DBLP:journals/cacm/CasadoFG14}.
The nodes of the network are \emph{switches} which forward \emph{packets} along the edges of the network according to the \emph{routing configuration}.
Packets enter the network at \emph{ingress switches} and leave it at \emph{egress switches}.
SDNs separate the  packet forwarding  process,  called  the \emph{data  plane}, from  the  routing process, called the \emph{control plane}.
\emph{Concurrent updates} to the routing configuration are difficult to get right~\cite{DBLP:conf/networking/ForsterMW16}.

The separation of data and control plane and updates to the routing configuration can be encoded into Petri nets with transits~\cite{DBLP:conf/atva/FinkbeinerGHO19}.
Using this encoding, we demonstrate the workflow of the web interface for model checking an asynchronous distributed system with data flows.
The packets of the SDN are modeled by the data flow in the Petri net with transits.
The data flow relation as an extension from Petri nets to Petri nets with transits is depicted as colored and labeled arcs.
In \refFig{pnwt}, the web interface presents the resulting Petri net with transits \(\pNet\).
First, we use the tools on the left to create for each switch a place \(si\) with \(i\in\{0,\ldots,5\}\) and add a token (cf.\ outer parts of \(\pNet\)).
Then, we create transitions for the connections between the switches and for the origin of packets in the SDN (cf.\ transition \(\mathit{ingress}\) in the top-left corner) and link them with flows in both directions.
Additionally, we create local transits between the switches corresponding to the forwarding of packets.
They are displayed in light blue and red and are identified by the letters.
This constitutes the \emph{data plane}.

Next, we define the \emph{control plane}, i.e., which forwarding is activated.
Each transition to forward packets is connected to a place \(ai\) with \(i\in\{0,\ldots,5\}\)
which has a token when the forwarding is configured initially (cf.\ places \(a3\), \(a4\), and \(a5\)) and no token otherwise (cf.\ places \(a0\), \(a1\), and \(a2\)).
For the concurrent update, we create places \(ui\) with \(i\in\{0,\ldots,7\}\) and transitions \(ti\) with \(i\in\{6,\ldots,11\}\) with corresponding flows (cf.\ inner parts of \(\pNet\)).

Transitions for the forwarding are set as weak fair, i.e., whenever a transition is infinitely long enabled in a run, it also has to fire infinitely often, indicated by the purple color of the outer transitions.
Transitions for the update do not require fairness assumptions.
A satisfied Flow-LTL formula is $A\,F\,s5 $ specifying that all packets eventually reach switch $s5$.
An unsatisfied formula is \((G\,u0\Rightarrow A\,F\,s2)\) requiring for runs, where the update is never executed, that all packets are taking the lower-left route.
The fairness assumptions and a maximality assumption, i.e., whenever some transition can fire in a run some transition fires, are automatically added to the formula.
In the screenshot, a counterexample for the unsatisfied formula is displayed on the right.
The first packet takes the upper-right route via transitions $t3$, $t4$, and $t5$ and the update never starts.

\subsubsection{Features for Petri Nets with Transits.}
\label{sec:PNwTgeneral}

\AdamMC~\cite{DBLP:conf/cav/FinkbeinerGHO20} is a command-line model checking tool for Petri nets with transits and Flow-LTL~\cite{DBLP:conf/atva/FinkbeinerGHO19}.
The model checking problem of Petri nets with transits against Flow-LTL is solved by a reduction to Petri nets and LTL.
The web interface allows displaying and arranging the nodes of the Petri net from the reduction and the input Petri net with transits.
Automatic layout techniques
are applied to avoid the overlapping of nodes.
A physics control, which modifies the repulsion, link, and gravity strength of nodes, can be used to minimize the overlapping of edges.
Heuristics generate coordinates for the constructed Petri net by using the coordinates of the input Petri net with transits to obtain a similar layout of corresponding parts.

For a positive result, the web interface allows visualizing the data flow trees for given firing sequences of the nets.
For a negative result, the counterexample can be simulated both in the Petri net with transits and in the Petri net from the reduction.
The witness of the counterexample for each flow subformula
and the run violating the global behavior
can be displayed by the web interface.
This functionality is helpful when developing an encoding of a problem into Petri net with transits
to ensure that a counterexample is not an error in the encoding.
The constructed Petri net can be exported into a standard format for Petri net model checking (PNML)
and the constructed LTL formula can be displayed.

\section{Web Interface for Petri Games}

The web interface can synthesize local controllers from safety specifications.
The workflow is showcased for a distributed alarm system given as a Petri game.

\subsubsection{Workflow for Petri Games}
\label{sec:PGworkflow}
\begin{figure}[t]
	\centering
	\includegraphics[width=\textwidth]{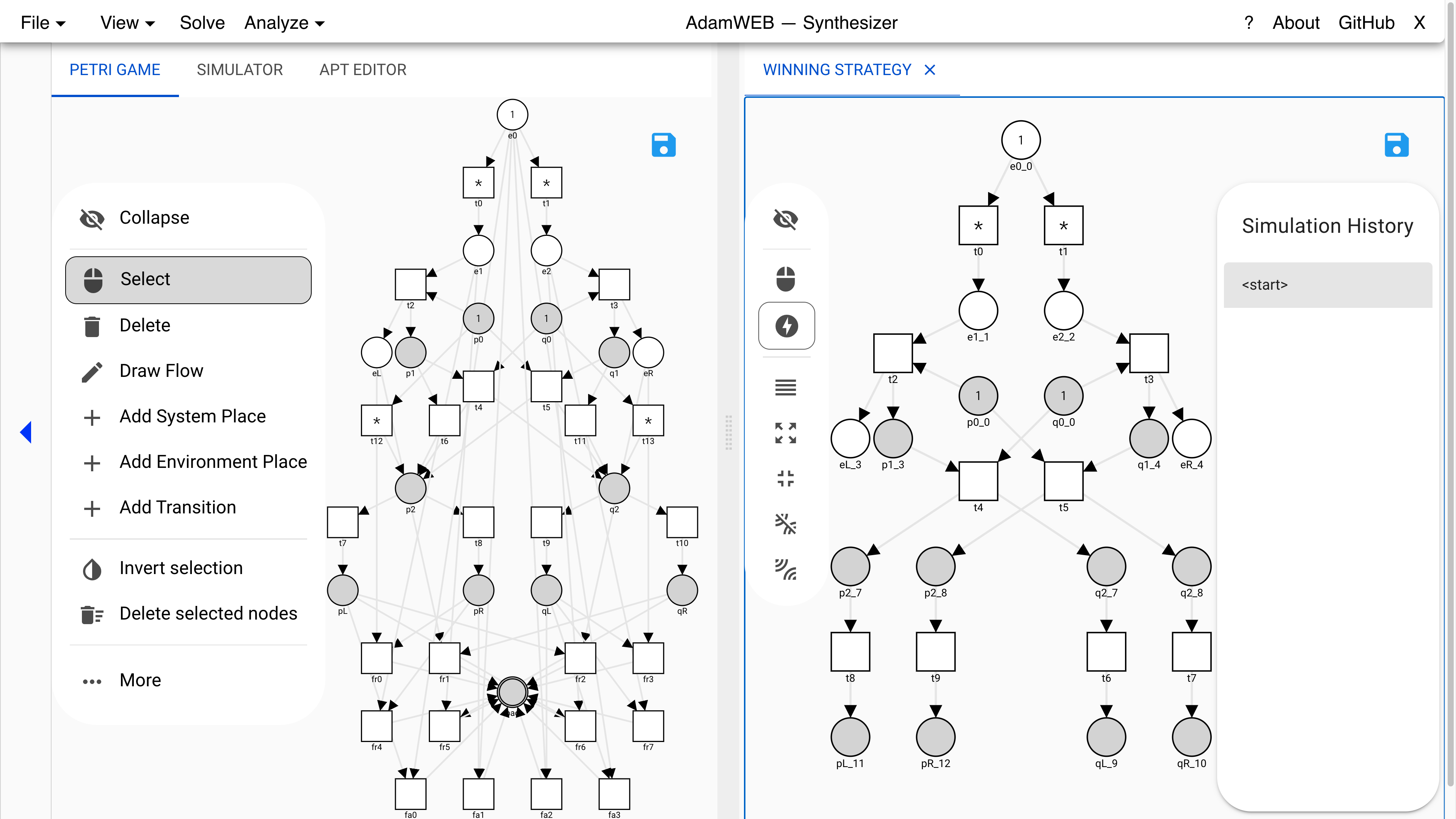}
	\caption{Screenshot from the web interface for the synthesis workflow.}
	\label{fig:pg}
\end{figure}
We demonstrate the workflow of the web interface for the synthesis of asynchronous distributed systems with causal memory from safety specifications.
Petri games separate the places of an underlying Petri net into \emph{system places} and \emph{environment places}.
Tokens on system places are \emph{system players} and tokens on environment places are \emph{environment players}.
Each player has \emph{causal memory}: only upon synchronization with other players, they exchange their entire causal past.
For safety specifications, the system players have to avoid that a bad place is reached for all behaviors of the environment players.

We want to obtain two local controllers of a distributed alarm system that should indicate the location of a burglary at both controllers.
In \refFig{pg}, the web interface presents the resulting Petri game on the left and the winning strategy for the alarm system on the right.
The burglar is modeled by an environment player and each component of the distributed alarm system by a system player.
Environment players are on white places and system players on gray ones.
We create five environment places $e0$, $e1$, $e2$, $\mathit{eL}$, and $\mathit{eR}$.
The place $e0$ has a token, $e1$ and $e2$ serve for the decision to burgle a location, and $\mathit{eL}$ and $\mathit{eR}$ for actually burgling the location.
Each component \(x\in\{p,q\}\) of the alarm system has one system place $x0$ with a token, two system places $x1$ and $x2$ to detect a burglary and inform the other component, and two system places $\mathit{xL}$ and $\mathit{xR}$ to sound an alarm with the position of a burglary.
We create rows of transitions for the environment player deciding where to burgle (first row), for the components detecting a burglary (second row), for the communication between the components (third row), and for sounding the alarm at each location (fourth row).

At last, we use transitions $\mathit{fa}i$ with $i\in\{0,\ldots,3\}$ and $\mathit{fr}j$ with $j\in\{0,\ldots,7\}$ connected to the bad place $\mathit{bad}$ to define that the implementation of the distributed alarm system should avoid false alarms and false reports.
A \emph{false alarm} occurs if the burglar did not burgle any location but an alarm occurred, i.e., in every pair of places $\{e0\}\times\{\mathit{pL}, \mathit{pR}, \mathit{qL}, \mathit{qR}\}$.
A \emph{false report} occurs if a burglary happened at a location but a component of the alarm system indicates a burglary at the other location, i.e., in every pair of places $\{e1, \mathit{eL}\} \times \{\mathit{pR}, \mathit{qR}\}$ and $\{e2, \mathit{eR}\} \times \{\mathit{pL}, \mathit{qL}\}$.
We add transitions and flows to $\mathit{bad}$ for these cases.

The web interface finds a winning strategy (depicted on the right in \refFig{pg}) for the Petri game described above.
Each component locally monitors its location ($t2$, $t3$) and simultaneously waits for information about a burglary at the other location ($t4$, $t5$).
When a burglary is detected at the location of the component then it first informs the other component ($t4$, $t5$) and then outputs an alarm for the current location ($t7$, $t8$).
When a component is informed about a burglary at the other location, it outputs an alarm for the other location ($t6$, $t9$).

\subsubsection{Features for Petri Games}
\label{sec:PGgeneral}

\AdamSYNT~\cite{DBLP:conf/cav/FinkbeinerGO15} is a command-line tool for Petri games~\cite{DBLP:journals/iandc/FinkbeinerO17}.
The synthesis problem for Petri games with a bounded number of system players, one environment player, and a safety objective is reduced to the synthesis problem for two-player games.
A winning strategy in the two-player game is translated into a winning strategy for the Petri game.
Both can be visualized in the web interface.
Here, the web interface provides the same features for visualizing, manipulating, and automatically
laying out the elements as for model checking.
It uses the order of nodes of the Petri game to heuristically provide a positioning of the strategy
and allows simulating runs of the strategy.
The winning strategy of the two-player game provides
an additional view on the implementation to check if it is not bogus
due to a forgotten case in the Petri game or specification.
For an unrealizable synthesis problem, the web interface allows analyzing
the underlying two-player game via a stepwise creation of strategies.
This guides the user towards changes to make the problem realizable.

\section{Implementation Details}
\label{sec:impldetails}
The server is implemented using the Sparkjava micro-framework~\cite{sparkjava} for incoming HTTP and WebSocket connections.
The client is a single-page application written in Javascript using Vue.js~\cite{vue}, D3~\cite{d3}, and the Vuetify component library~\cite{vuetify}.
We constructed libraries out of the tools \AdamMC{} and \AdamSYNT{} and implemented one interface handling both libraries.
Common features like the physics control of nodes share the same implementation.
All components of the libraries and the web interface~\cite{webinterface} are open source and available on GitHub~\cite{adamtool}.

\section{Conclusion}
\label{sec:conclusion}
We presented a web interface for two tools:
\AdamMC{}, a model checker for data flows in asynchronous distributed systems represented by Petri nets with transits, and
\AdamSYNT{}, a synthesis tool for local controllers from safety specifications in asynchronous distributed systems with causal memory represented by Petri games.
The web interface makes the modeling and debugging of Petri nets with transits and Petri games user-friendly as it presents visual representations of the input, all intermediate steps, and the output of the tools.
The interactive features are a great assistance for correctly modeling distributed systems.

We plan to extend the web interface and tool support to model checking Petri nets with transits against Flow-CTL$^*$~\cite{DBLP:conf/atva/FinkbeinerGHO20}, to other classes of Petri games with a decidable synthesis problem~\cite{DBLP:conf/fsttcs/FinkbeinerG17,DBLP:conf/concur/BeutnerFH19}, to the bounded synthesis approach for Petri games~\cite{DBLP:conf/birthday/Finkbeiner15,DBLP:journals/corr/abs-1711-10637,DBLP:journals/corr/Tentrup16,DBLP:conf/atva/Hecking-Harbusch19}, and to high-level Petri games~\cite{DBLP:journals/acta/GiesekingOW20}.
As our web interface is open source and easy to extend, we also plan to connect it to other tools for Petri nets like APT~\cite{apt}, LoLA~\cite{DBLP:conf/apn/Wolf18a}, or TAPAAL~\cite{DBLP:conf/tacas/DavidJJJMS12}.

\bibliographystyle{splncs04}
\bibliography{bib}
\end{document}